\def\Ar{\rightarrow}

\def\b{\beta}
\def\n{\nu}

\def\e{\epsilon}

\def\t{\tilde}

\def\G{{\rm GeV}}

\def\eV{{\rm eV}}
\documentstyle [12pt,epsf]{article}
\begin{document}
\baselineskip=24pt
\setcounter{page}{1}
\thispagestyle{empty}
\topskip 1.5  cm
\topskip 0.5  cm
\begin{flushright}
\begin{tabular}{c c}
& {\normalsize hep-ph/0110283}\\
\end{tabular}
\end{flushright}
\vspace{1 cm}
\centerline{\LARGE \bf Lepton Flavor Violating Process in  Degenerate}
\vskip 0.5 cm
\centerline{\LARGE \bf and Inverse-Hierarchical Neutrino Models}
\vskip 1.5 cm
\centerline{{\large \bf Atsushi Kageyama}
 \renewcommand{\thefootnote}{\fnsymbol{footnote}}
\footnote[1]{E-mail address:  atsushi@muse.hep.sc.niigata-u.ac.jp},
\qquad{\large \bf Satoru Kaneko}
\renewcommand{\thefootnote}{\fnsymbol{footnote}}
\footnote[2]{E-mail address: kaneko@muse.hep.sc.niigata-u.ac.jp}}
\vskip 0.8 cm
\centerline{{\large \bf Noriyuki Shimoyama}
\renewcommand{\thefootnote}{\fnsymbol{footnote}}
\footnote[3]{E-mail address: simoyama@muse.hep.sc.niigata-u.ac.jp},
\qquad {\large \bf Morimitsu Tanimoto}
\renewcommand{\thefootnote}{\fnsymbol{footnote}}
\footnote[4]{E-mail address: tanimoto@muse.hep.sc.niigata-u.ac.jp}
 }
\vskip 0.8 cm
 \centerline{ \it{Department of Physics, Niigata University, 
 Ikarashi 2-8050, 950-2181 Niigata, JAPAN}}
\vskip 2.2cm
\centerline{\bf ABSTRACT}\par
\vskip 0.5 cm
We have investigated the lepton flavor violation in the supersymmetric 
framework assuming the large mixing angle MSW solution with the 
quasi-degenerate and the inverse-hierarchical neutrino masses.
In the case 
of the quasi-degenerate neutrinos, the predicted branching ratio
BR$(\mu \rightarrow e \gamma)$ strongly depends on $m_\nu$ and $U_{e3}$.
For $U_{e3}\simeq 0.05$ with $m_\nu \simeq 0.3\ \eV$, the prediction is close
to the present experimental upper bound if the right-handed Majorana 
neutrino masses are degenerate.  On the other hand, the prediction is
larger than the experimental upper bound for $U_{e3}\geq 0.05$ in the case of
the inverse-hierarchical neutrino masses.
\newpage
\topskip 0. cm

 Super-Kamiokande  has almost confirmed  the neutrino oscillation
 in the atmospheric neutrinos, which favors the $\n_\mu\Ar \nu_\tau$
process  \cite{SKam}.
For the solar neutrinos \cite{SKamsolar,SNO}, the recent data of 
the Super-Kamiokande and the SNO also favor the neutrino oscillation
$\n_\mu\Ar \nu_e$ with the large mixing angle(LMA) MSW solution 
\cite{MSW,Lisi}. 
These  results  mean that neutrinos are  massive, moreover,
they indicate the bi-large flavor mixing in the lepton sector. 
The non-zero neutrino masses clearly imply new physics 
beyond the standard model (SM), and the large flavor mixings suggest
 that the flavor structure in the lepton sector is  very different 
from that in  the quark sector. 
 
If neutrinos are massive and mixed in the SM, 
there exists a source of the lepton flavor violation (LFV).   However, due to the smallness of the neutrino masses, 
the predicted branching ratios for these processes are  tiny \cite{SM}
 such as  BR$(\mu \rightarrow e \gamma) < 10^{-50}$.

On the other hand,
in the supersymmetric framework  the situation is completely different.
The SUSY  provides new direct sources of flavor violation in the lepton
 sector, namely the possible presence of off-diagonal soft terms
in the slepton mass matrices $\left(m_{\tilde L}^2\right)_{ij}$,
$\left(m_{\t e_R}^2\right)_{ij}$,  and trilinear couplings $A^e_{ij}$.
Strong bounds on these matrix elements come from requiring branching ratios 
for LFV processes to be below the observed ratios.  
For the present the strongest bound  comes from  the 
$\mu \rightarrow e \gamma $ decay.

In order to avoid these dangerous off-diagonal terms, one often 
imposes the  perfect universality of the $\left(m_{\t L}^2\right)_{ij}$,
$\left(m_{\t e_R}^2\right)_{ij}$,  $A^e_{ij}$ matrices, i.e. to take them
proportional to the unit matrix.   
However, even under the universality assumption, 
 radiative corrections generate off-diagonal soft terms 
due to the massive neutrinos.
The flavor
changing operators giving rise to the non-diagonal  neutrino mass matrices
will  contribute to the renormalization group equations (RGE's)
of the $\left(m_{\t L}^2\right)_{ij}$, $\left(m_{\t e_R}^2\right)_{ij}$ and
$A^e_{ij}$ matrices and  induces  off-diagonal entries.

Suppose that  neutrino masses are produced by the  see-saw mechanism
 \cite{seesaw}, there are  the right-handed neutrinos
 above a  scale $M_R$.  Then neutrinos 
have  the Yukawa coupling matrix $Y_\nu$  with  off-diagonal entries
in the basis of the diagonal charged-lepton Yukawa couplings.
 The  off-diagonal entries of $Y_\nu$ drive off-diagonal entries 
in the $\left(m_{\t L}^2\right)_{ij}$ and $A^e_{ij}$ matrices 
through the RGE's running \cite{Borzumati}.

Some authors \cite{LFV1,LFV2,Sato,Casas} 
have already  analyzed the effect of neutrino Yukawa couplings
for the LFV  focusing on the recent data of the Super-Kamiokande.
Those analyses have been done assuming that neutrino masses
are  hierarchical ones, which is  similar to  quarks and charged-leptons.
However, since the data of neutrino oscillations only indicate
the differences of the mass square   $\Delta m^2_{ij}$,
the neutrinos may have  the quasi-degenerate spectrum 
 $m_1\simeq m_2\simeq m_3$
or the  inverse-hierarchical one  $m_1\simeq m_2 \gg  m_3$.
These cases have been  discussed qualitatively neglecting
$U_{e3}$ in ref. \cite{Casas}.
In our work, we present quantitative analyses of  
BR($\mu \rightarrow e \gamma $) including effect of $U_{e3}$  in the case of
 the quasi-degenerate and inverse-hierarchical  neutrino masses.
It should be emphasized that  the magnitude of  $U_{e3}$ is one of important
ingredients to predict the branching ratio.

 In terms of the standard parametrization of the mixing matrix  \cite{PDG},
 the MNS matrix $U$ (lepton mixing matrix) \cite{MNS} are given as

\begin{equation}  
 U =  \left (\matrix{ c_{13} c_{12} & c_{13} s_{12} &  s_{13} e^{-i \phi}\cr 
  -c_{23}s_{12}-s_{23}s_{13}c_{12}e^{i \phi} & 
c_{23}c_{12}-s_{23}s_{13}s_{12}e^{i \phi} &   s_{23}c_{13} \cr
  s_{23}s_{12}-c_{23}s_{13}c_{12}e^{i \phi} & 
 -s_{23}c_{12}-c_{23}s_{13}s_{12}e^{i \phi} &  c_{23}c_{13} \cr} \right ) \ ,
\end{equation} 

\noindent  where  $s_{ij}\equiv \sin{\theta_{ij}}$ and 
 $c_{ij}\equiv \cos{\theta_{ij}}$ are mixings in vacuum, 
 and $\phi$ is the $CP$ violating phase.   
Assuming that oscillations need only accounting for 
 the solar and the atmospheric neutrino data, we take  the 
LMA-MSW solution, in which 
the magnitude of the  MNS matrix elements are given in ref. \cite{FT} 
 and the neutrino mass scales are given by the data
\begin{equation}
\Delta m^2_{\rm atm}=  (1.5\sim 5)\times  10^{-3} \eV^2\ , \quad
\Delta m_{\odot}^2= 2.5\times  10^{-5}\sim 1.6\times  10^{-4}\eV^2 \ . 
\end{equation}

In our calculation of the LFV effect, we take the typical values 
 $\Delta m^2_{\rm atm}=  3\times  10^{-3} \eV^2 , \
\Delta m_{\odot}^2= 7 \times  10^{-5}  \eV^2$ 
and $s_{23}=1/\sqrt{2}, \ s_{12}=0.6, \ s_{13}\leq 0.2$,
 taking account of the CHOOZ  data \cite{Chooz}.  
The CP violating phase is neglected for simplicity.

First, let us consider the degenerate neutrino case.
The neutrino masses are given as
\begin{equation}
 m_1\equiv m_\nu \ ,  
 \qquad  m_2=m_\nu + \frac{1}{2 m_\nu}\Delta m^2_{\odot}\  , \qquad
 m_3=m_\nu +  \frac{1}{2 m_\nu} \Delta m_{\rm atm}^2 \ , 
\end{equation}
\noindent in terms of the experimantal values
 $\Delta m_{\rm atm}^2$ and  $\Delta m^2_{\odot}$. 
Since the best bound on the neutrinoless double beta decay
obtained by Heidelberg-Moscow group gives \cite{Beta}
$m_{ee}< 0.34\  \eV$ at 90\% C.L., we take $m_{\nu}=0.3\ \eV$ 
in the following calculations.  The $m_{\nu}$ dependence of our result is 
commented later.\footnote{If we take account of the non-zero Majorana phases,
we can take  $m_{\nu}$ larger than $0.34\ \eV$.}

The see-saw mechanism leads to the tiny neutrino mass
as follows $Y_\nu^T M_R^{-1} Y_\nu v^2$, where 
$v$ is the vacuum expectation value
of the relevant Higgs field and  $M_R$ is the right-handed Majorana
neutrino mass matrix. Then, we can get the Yukawa coupling 
 in the basis of the  diagonal charged-lepton Yukawa couplings as \cite{Casas}

\begin{equation}
Y_\nu= \frac{1}{v}  \left (\matrix{\sqrt{M_{R1}}& 0 & 0\cr
  0 & \sqrt{M_{R2}}  & 0 \cr  0 & 0 & \sqrt{M_{R3}} \cr  } \right) 
  R
  \left (\matrix{\sqrt{m_1}& 0 & 0\cr
  0 & \sqrt{m_2}  & 0 \cr  0 & 0 & \sqrt{m_3} \cr  } \right)  
 U^\dagger \ ,
\label{YEW}
\end{equation}
where $R$ is an orthogonal $3\times 3$ matrix and depends on the model.
We take $R$ to be  the unit matrix and  degenerate right-handed Majorana
masses  $M_{R1}=M_{R2}=M_{R3}\equiv M_R$.  
This assumption is derived logically, otherwise a big conspiracy would be 
needed between $Y_\nu$ and $M_R$.  Actually,  this assumption is realized
 in some models as discussed later.
However,  as the  conspiracy between $Y_\nu$ and $M_R$ is not excluded
 a priori, we also discuss the effect of the non-degenerate right-handed
Majonara neutrino masses.

Eq.(\ref{YEW}) presents the Yukawa coupling at the electroweak scale.
Since we need the Yukawa coupling at the GUT scale or
the Planck scale, 
 eq.(\ref{YEW}) should be modified by taking account of the effect of 
the RGE's \cite{RGE1,RGE2,Haba}.
Modified Yukawa couplings at a scale $M_R$   are given as 
\begin{equation}
Y_\nu= \frac{\sqrt{M_R}}{v}  \left (\matrix{\sqrt{m_1}& 0 & 0\cr
  0 & \sqrt{m_2}  & 0 \cr  0 & 0 & \sqrt{m_3} \cr  } \right)  
 U^\dagger \sqrt{I_g I_t}
\left (\matrix{1& 0 & 0\cr 0 & 1 & 0 \cr 0 & 0 & \sqrt{I_\tau}\cr}\right)\ ,
\label{YGUT}
\end{equation}
\noindent
with
\begin{equation}
I_g =\exp\left [\frac{1}{8\pi^2}\int_{t_Z}^{t_R} -c_i g_i^2 dt\right ]\ ,\quad
I_t =\exp \left [\frac{1}{8\pi^2}\int_{t_Z}^{t_R} y_t^2 dt \right ] \ ,  \quad
I_{\tau}= \exp\left [\frac 1{8\pi^2}\int_{t_Z}^{t_R} y_{\tau}^2 dt \right ]\ ,
\end{equation}
\noindent where  $t_R=\ln M_R$ and  $t_Z=\ln M_Z$. Here, $g_i(i=1-2)$'s are 
gauge couplings and  $y_t$ and $y_\tau$ are Yukawa couplings. 

  In practice, the degenerate neutrino masses with the LMA-MSW solution are 
predicted by the models with the flavor symmetry  $S_{3R}\times S_{3L}$ 
\cite{FTY} or  $O_{3R}\times O_{3L}$ \cite{O3},\footnote{These models 
predicted the effective left-handed Majorana neutrino
 masses by the symmetry without introducing the see-saw mechanism.
However, one can easily get the see-saw realization by introducing
the right-handed neutrinos  $\nu_R$.  Assuming the same transformation 
property of  the flavor symmetry  for both $\nu_L$ and  $\nu_R$, 
one obtains $M_R\propto {\rm I}$ and $Y_\nu\propto {\rm I}$.}
which gives the democratic  mass matrices in the quark sector \cite{Demo}.
In those models, $U_{e3}$ is predicted to be $\sim 0.05$.
We will show  results for both cases of $U_{e3}=0.05$ and  $0.2$. 


Second, let us consider the case of the inverse-hierarchical neutrino masses. The neutrino masses are given as
\begin{equation}
   m_2\equiv \sqrt{\Delta m^2_{\rm atm}} \  , 
\quad m_1=m_2-\frac{1}{2 m_2}\Delta m_{\odot}^2 \ ,  
 \quad   m_3\simeq 0 \ ,
\end{equation}
\noindent where $m_2>m_1$ is taken in order to keep the MSW effect.

The typical model of the inverse-hierarchy is the Zee model \cite{Zee},
in which the right-handed neutrinos do not exist.
However, one can also consider Yukawa textures  which lead to the inverse mass 
hierarchy through the see-saw mechanism as seen in eq.(\ref{YEW}).
The detailed analyses with definite models \cite{Shafi} will show 
in a further comming paper.
In this work, we take $R=I$   and  degenerate right-handed 
Majorana masses  $M_{R1}=M_{R2}=M_{R3}\equiv M_R$
as well as the case of degenerate neutrino masses.  
 We  also discuss the case of the non-degenerate right-handed
Majonara neutrino masses.

In the presence of non-zero neutrino Yukawa couplings, we can
expect the LFV phenomena in the lepton sector.
Within the framework of SUSY models, 
the flavor violation in neutrino Yukawa couplings induces 
the LFV in slepton masses even if we assume the universal scalar mass for 
all scalars at the GUT scale.
In the present models, the LFV is generated
in left-handed slepton masses since right-handed neutrinos couple to 
the left-handed lepton multiplets. A RGE  
for the left-handed doublet slepton masses $(m^2_{\tilde{L}})$ can be 
written as

\begin{eqnarray}
\mu \frac{d(m^2_{\tilde{L}})_{ij}}{d\mu}
&=& \left(\mu \frac{d(m^2_{\tilde{L}})_{ij}}{d\mu} \right)_{\rm MSSM}
\nonumber \\
&&\hspace{-2cm}+\frac{1}{16\pi^2} \left[
m^2_{\tilde{L}} Y_\nu^\dagger Y_\nu+Y_\nu^\dagger Y_\nu m^2_{\tilde{L}}
+2(Y_\nu^\dagger m^2_{\tilde \nu} Y_\nu +\widetilde{m}^2_{H_u}
Y_\nu^\dagger Y_\nu +A_\nu^\dagger A_\nu)\right]_{ij},
\end{eqnarray}
where $m^2_{\tilde{\nu}}$ and $\widetilde{m}^2_{H_u}$ are soft SUSY-breaking 
masses for right-handed sneutrinos ($\tilde{\nu}$) and doublet Higgs
($H_u$), respectively.
Here $(\mu d(m^2_{\tilde{L}})_{ij}/{d\mu} )_{\rm MSSM}$ denotes the RGE
 in case of the minimal SUSY standard model (MSSM), and the terms explicitly 
written are additional 
contributions in the presence of the neutrino Yukawa couplings.
In a basis where the charged-lepton Yukawa couplings are diagonal, the term
$(\mu d(m^2_{\tilde{L}})_{ij}/{d\mu} )_{\rm MSSM}$ does not provide
any flavor violations. Therefore the only source of LFV comes 
from the additional terms. 
In our analysis, we numerically solve the RGE's, 
and then calculate the event rates for the LFV processes by using the 
complete formula in ref.~\cite{LFV1}. 
Here, in order to obtain an approximate estimation for the LFV masses,
let us consider  approximate solution to the LFV mass terms ($i\neq j$):
\begin{eqnarray}
(\Delta m^2_{\tilde{L}})_{ij} \simeq -\frac{(6+2 a_0^2)m_0^2}{16 \pi^2}
(Y_\nu^\dagger Y_\nu)_{ij} \ln\frac{M_X}{M_R}\ ,
\end{eqnarray}
where we assume a universal scalar mass $(m_0)$ for all scalars and
a universal A-term $(A_f=a_0 m_0 Y_f)$ at the GUT scale $M_X$. 
It is noticed that large neutrino Yukawa couplings and large lepton mixings 
 generate the large LFV in the left-handed slepton masses.

The decay rates can be approximated as follows:
\begin{eqnarray}
{\rm \Gamma} (e_i \rightarrow e_j \gamma) \simeq 
\frac{e^2}{16 \pi} m^5_{e_i} F\left| (\Delta m^2_{\tilde{L}})_{ij}\right|^2\ ,
\label{event_rate}
\end{eqnarray}
where  $F$ is a function of masses and mixings for SUSY particles.

Before showing  numerical results, we present a qualitative discussion
on $(Y_{\nu}^{\dag}Y_{\nu})_{21}$, which is a crucial quantity
to predict the branching ratio BR$(\mu \rightarrow e \gamma)$.   
This is given in terms of neutrino masses and mixings at the electroweak scale
 as follows:
\begin{equation}
(Y_{\nu}^{\dag}Y_{\nu})_{21}=
\frac{M_{\rm R}}{v_{\rm u}^{2}}
\left[
U_{\mu2}U_{e2}^{*}(m_{2}-m_{1})+U_{\mu3}U_{e3}^{*}(m_{3}-m_{1})
\right] \ ,
\label{ff}
\end{equation}
\noindent   where $v_u\equiv {\rm v}\sin\b$ with ${\rm v}=174\G$ 
is taken as an usual notation
 and a unitarity relation of the MNS matrix elements is used.
Taking the three cases of the neutrino mass spectra, the degenerate, 
the inverse-hierarchical  and the normal hierarchical masses,
one obtains
\begin{eqnarray}
(Y_{\nu}^{\dag}Y_{\nu})_{21}\simeq &&
\frac{M_{\rm R}}{2\sqrt{2}v_{\rm u}^{2}}\frac{\Delta m^2_{\rm atm}}{m_\nu}
\left[\frac{1}{\sqrt{2}}U_{e2}^* 
 \frac{\Delta m_{\odot}^2}{\Delta m^2_{\rm atm}}+ U_{e3}^*\right] \ ,          \quad ({\rm Degenerate}) \nonumber \\
 \nonumber \\
\simeq  &&
\frac{M_{\rm R}}{\sqrt{2}v_{\rm u}^{2}}\sqrt{\Delta m_{\rm atm}^2}
\left[ \frac{1}{2\sqrt{2}} U_{e2}^*
\frac{\Delta m_{\odot}^2}{\Delta m^2_{\rm atm}} - U_{e3}^*\right] \ , 
         \quad ({\rm Inverse }) 
 \label{mass3} \\
 \nonumber \\
\simeq  &&
\frac{M_{\rm R}}{\sqrt{2}v_{\rm u}^{2}}\sqrt{\Delta m_{\rm atm}^2}
\left[\frac{1}{\sqrt{2}}U_{e2}^*
 \sqrt{\frac{\Delta m_{\odot}^2}{\Delta m^2_{\rm atm}}} + U_{e3}^*\right] \ , 
         \quad ({\rm Hierarchy }) \nonumber
\end{eqnarray}
\noindent
where we take the maximal mixing for the atmospheric neutrinos. 
Since  $U_{e2}\simeq 1/\sqrt{2}$ for the bi-maximal mixing matrix, 
the first terms in the square brackets of the right hand sides of  
eqs.(\ref{mass3}) can be  estimated by putting the experimental data.
We should take care the magnitude of  $U_{e3}$ in the second terms
to predict  the branching ratios 
because the second term is the dominant one as far as $U_{e3}\geq 0.01\ (0.1)$
 for the degenerate and the inverse-hierarchical  cases 
(for the normal hierarchical case).
For the case of the degenerate neutrino masses,
$(Y_{\nu}^{\dag}Y_{\nu})_{21}$ depends on the unkown neutrino mass scale 
$m_\nu$.  As one takes the larger $m_\nu$, one predicts
 the smaller branching ratio.  In our calculation, we take 
 $m_{\nu}=0.3 \ \eV$,  which  is close to 
the upper bound of the neutrinoless double beta decay experiment. 

As seen in  eqs. (\ref{mass3}),
 we expect that the case of the degenerate neutrino masses gives
the smallest branching ratio  BR$(\mu \rightarrow e \gamma)$.
The comparison in  cases of the inverse-hierarchical masses and the normal
hierarchical masses depends on the magnitude and phase of  $U_{e3}$.
In the limit of  $U_{e3}=0$, the case of the normal hierarchical masses
predicts  larger branching ratio. However, for $U_{e3}\simeq 0.2$ 
the predicted branching ratios are almost same in both cases.


   Let us present numerical calculations in the case of the degenerate 
neutrino masses 
assuming a universal scalar mass $(m_0)$ for all scalars and
 $a_0=0$ as a universal A-term at the GUT scale ($M_X=2\times 10^{16}$ GeV).
 We show  BR$(\mu \rightarrow e \gamma)$ versus the left-handed 
selectron mass $m_{\tilde e_L}$ for each $\tan\b=3,\  10,\  30$
and a fixed wino mass $M_2$ at the electroweak scale.
In fig.1, the branching ratio is shown for  $M_2=150,\  300\ \G$
 in the case of $U_{e3}=0.2$ with $M_R=10^{14} \G$,
in which the solid curves correspond to  $M_2=150 \G$
and the dashed ones to  $M_2=300 \G$.
The threshold of the selectron mass is determined by the
recent LEP2 data \cite{LEP} for   $M_2=150 \G$, but determined
by the  constraint that  the left-handed slepton  should be 
 heavier than the neutralinos for  $M_2=300 \G$.
As the  $\tan\b$ increases, the branching ratio increases
because the decay amplitude from the SUSY diagrams is approximately 
proportional to  $\tan\b$ \cite{LFV1}.
It is found that the branching ratio is  almost larger than the
experimental upper bound in the case of   $M_2=150 \G$.
On the other hand, the predicted values are smaller than the
experimental bound except for $\tan\b=30$  in the case of $M_2=300 \G$.

 In order to compare the degenerate neutrino mass case  with 
 the hierarchical one ($m_3\gg m_2 \gg m_1$), 
 we show the results of  the hierarchical case in fig.2, 
 in which parameters are taken as same as in fig.1.
It is remarked that the branching ratio is suppressed
 in the degenerate case.

Since the typical models \cite{FTY,O3} predict
$U_{e3}\simeq 0.05$, we also show the branching ratio for $U_{e3}=0.05$
in fig.3.  In this case, our predictions lie under the experimental upper 
bound even in the case of $M_2=150 \G$  except for $\tan\b=30$,
however, it is not far away from the experimental bound.
Therefore, we expect the observation of $\mu\Ar e\gamma$ in the near future.

 Our predictions depend on $M_R$ strongly, because the magnitude of
the neutrino Yukawa coupling is determined by  $M_R$
as seen in eq.(\ref{YGUT}).
If  $M_R$ reduces to  $M_R=10^{12} \G$,
the branching ratio reduces to $10^{-4}$ roughly  since it is proportional
to $M_R^2$.
The numerical result is shown in fig.4.
Thus, the branching ratio is not so large compared with the predictions
in the case of the  hierarchical neutrino masses \cite{Sato}.

It may be important to comment on the effect of the non-degenerate
right-handed Majorana neutrino masses since our results depend on 
the degeneracy of the right-handed Majorana neutrino masses.
We replace $(Y_{\nu}^{\dag}Y_{\nu})_{21}$  in eq.(\ref{ff}) with
\begin{equation}
(Y_{\nu}^{\dag}Y_{\nu})_{21}=
\frac{M_{\rm R}}{v_{\rm u}^{2}}
\left[
U_{\mu2}U_{e2}^{*}(\e_2 m_{2}-\e_1 m_{1})+U_{\mu3}U_{e3}^{*}(m_{3}-\e_1 m_{1})
\right] \ ,
\label{ffIn}
\end{equation}
\noindent where $\e_1=M_{R1}/M_{R3}$,  $\e_2=M_{R2}/M_{R3}$ 
and $M_{R}=M_{R3}$.
 The degenerate right-handed Majorana neutrino masses 
correspond to $\e_1=\e_2=1$.
If $\e_1$ and $\e_2$ deviate from 1 (the  non-degenerate case),
the cancellation among $\n_1$, $\n_2$ and $\n_3$ is weakened.
In other words,  the LFV is not so suppressed.
In  the case $\e_1 \ll \e_2 \ll 1$, the result is  similar to the one in
the case of hierarchical neutrino masses.


 Next we show the  results in the case of the inverse-hierarchical neutrino 
masses.  As expected in eq.(\ref{mass3}), the branching ratio
is much larger than the one in the degenerate case.
 In fig.5,
the branching ratio is shown for  $M_2=150,\  300\ \G$
 in the case of $U_{e3}=0.2$ with $M_R=10^{14} \G$.
 In fig.6, the branching ratio is shown for 
$U_{e3}=0.05$ with  $M_R=10^{14} \G$.
The  $M_R$ dependence is similar to the case of 
 the degenerate neutrino masses.
The predictions almost exceed the experimental bound  as far as
$U_{e3}\geq 0.05$, $\tan\beta\geq 10$ and  $M_R\simeq 10^{14} \G$.

These results also depend on the degeneracy of the right-handed Majorana 
neutrino masses.
As seen in eq.(\ref{ffIn}), the cancellation between $\n_1$ and $\n_2$
is weakened if $\e_1 \not = \e_2$. 
The predicted branching ratio depends on $\epsilon_{1}$ and
$\epsilon_{2}$.
For example, it could be larger than the one in the case of
hirarchical neutrino masses if   $\e_1 \ll \e_2 \simeq 1$ is realized in 
the right-handed Majorana neutrino mass matrix.
Actually, the typical model \cite{Shafi} 
does not guarantee our assumption  $M_R=I$, we need careful analyses,
which will be presented in a further comming paper.

\noindent{\bf Summary:}

 We have investigated the  LFV effect in the supersymmetric
framework  assuming LMA-MSW
 solution with the quasi-degenerate and the inverse-hierarchical case of 
neutrino masses.
 We show the predicted branching ratio of  $\mu \rightarrow e \gamma$
for both cases.
We expect the relation 
BR(degenerate) $\ll$ BR(inverse) $<$ BR(hierarchy) 
if the right-handed Majorana neutrino masses are degenerate.
In the case of the quasi-degenerate neutrinos, the predicted branching ratio
strongly depends on $M_R$, $m_\nu$ and $U_{e3}$. 
 For $U_{e3}\simeq 0.05$ with $m_\nu \simeq 0.3\ \eV$,
the prediction is close to the present  experimental upper bound.
On the other hand,  the prediction is  larger than the experimental 
upper bound for  $U_{e3}\geq 0.05$ 
 in the case of the inverse-hierarchical neutrino masses.
 More analyses including the process  $\tau \rightarrow \mu \gamma$
will be present elsewhere.

\vskip 1.5 cm
 
 We would like to thank Drs. J. Sato and H. Nakano for useful discussions.
 We also thank the organizers and participants of Summer 
Institute 2001 held at Yamanashi, Japan for helpful discussions. 
 This research is  supported by the Grant-in-Aid for Science Research,
 Ministry of Education, Science and Culture, Japan(No.10640274, No.12047220). 
\newpage

\newpage

\begin{figure}
\epsfxsize=15.0 cm
\vspace{-2.5 cm}
\centerline{\epsfbox{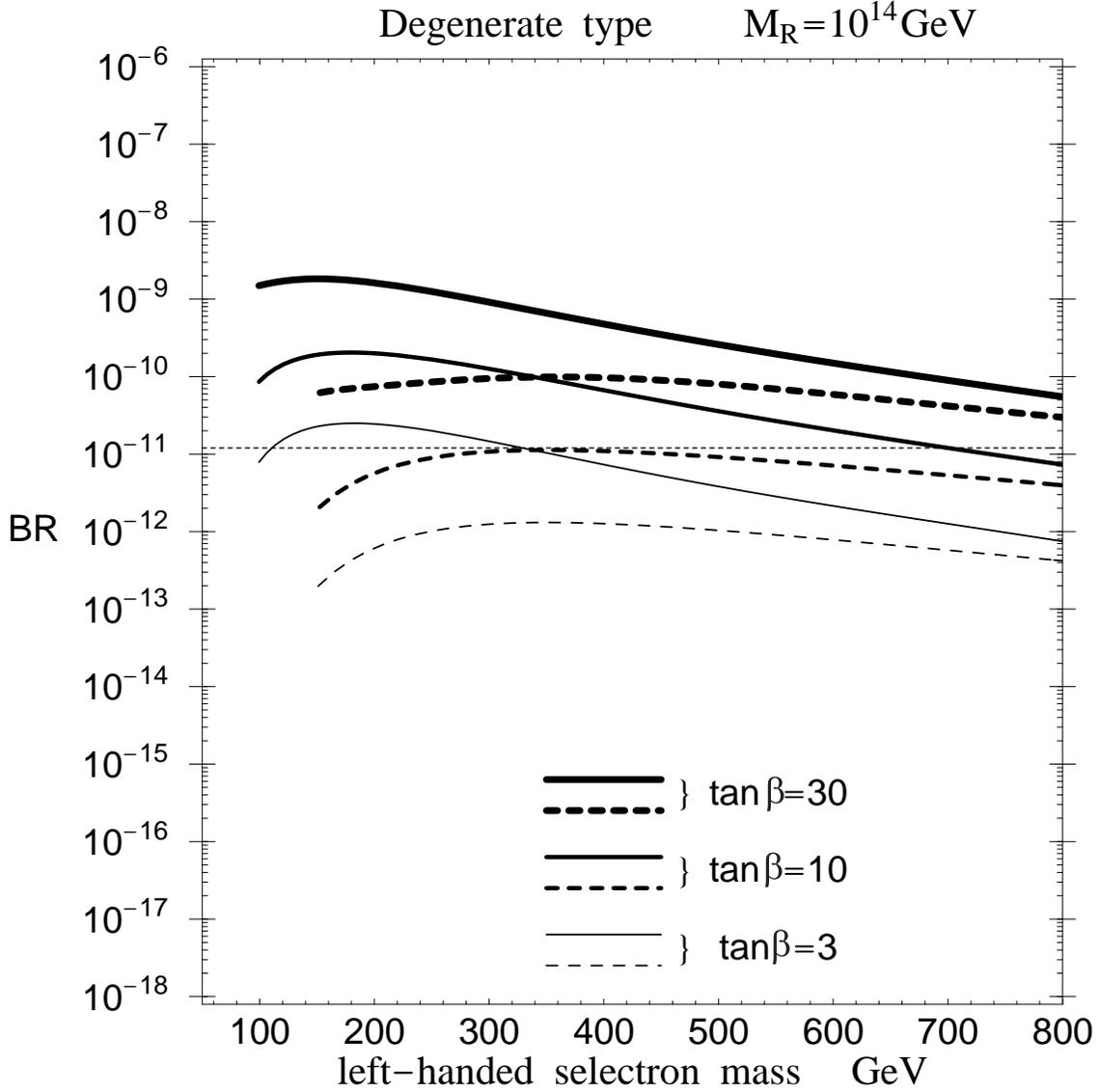}}
\caption{Predicted branching ratio BR$(\mu \rightarrow e \gamma)$
versus the left-handed selectron mass for $\tan\b=3,\ 10,\ 30$
in the case of the degenerate neutrino masses.
Here $M_R=10^{14}\G$ and $U_{e3}=0.2$ are taken.
The solid curves correspond to  $M_2=150 \G$ and the dashed ones to  
$M_2=300 \G$.
A horizontal dotted line denotes the experimental upper bound.}
\end{figure}

\begin{figure}
\epsfxsize=15.0 cm
\vspace{-2.5 cm}
\centerline{\epsfbox{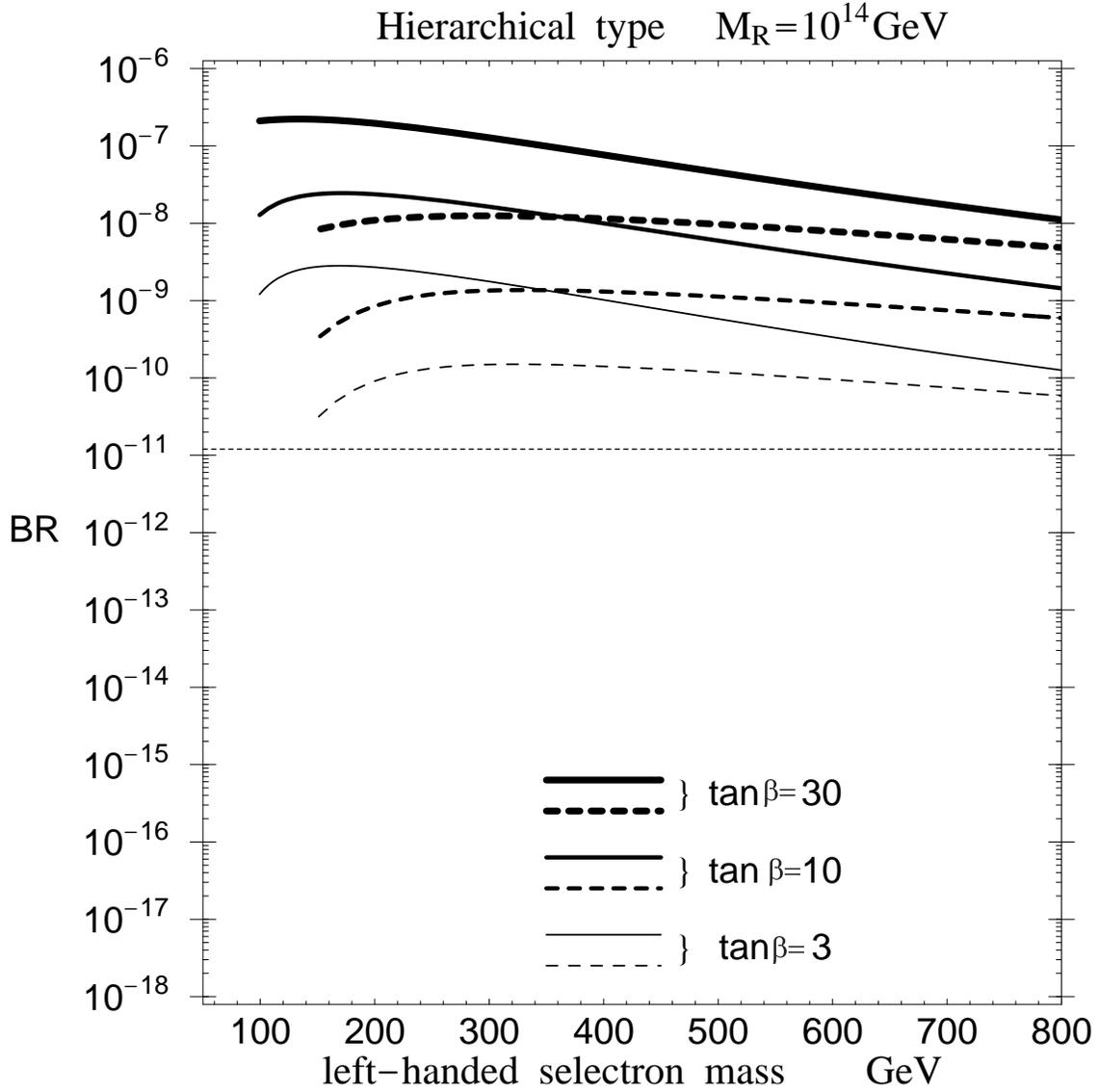}}
\caption{Predicted branching ratio BR$(\mu \rightarrow e \gamma)$
in the case of the hierarchical  neutrino masses.
Parameters are taken as same as in Fig.1.}
\end{figure}

\begin{figure}
\epsfxsize=15.0 cm
\vspace{-2.5 cm}
\centerline{\epsfbox{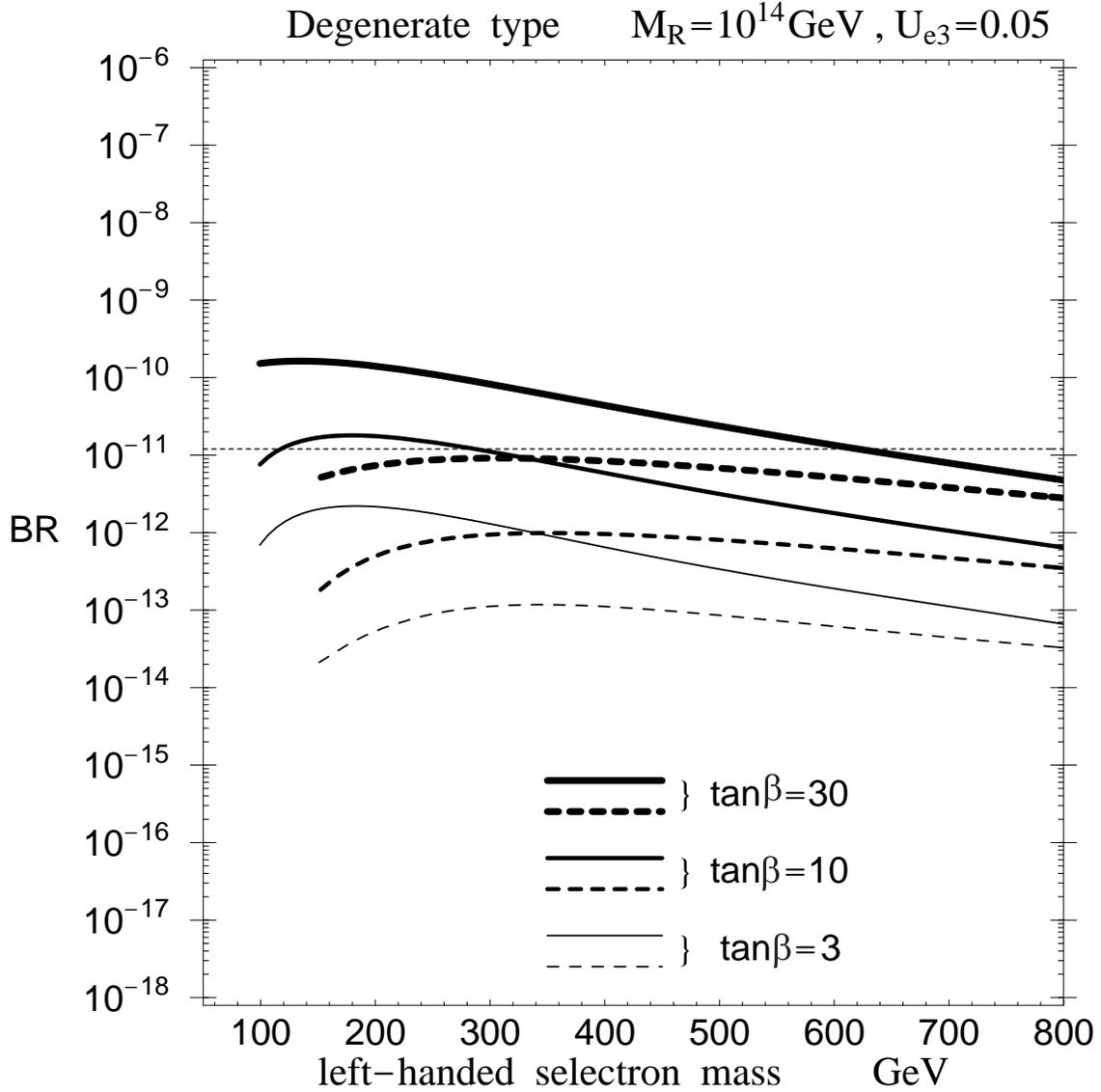}}
\caption{Predicted branching ratio BR$(\mu \rightarrow e \gamma)$
versus the left-handed selectron mass for $\tan\b=3,\ 10,\ 30$
in the case of the degenerate neutrino masses.
Here $M_R=10^{14}\G$ and $U_{e3}=0.05$ are taken.
The solid curves correspond to  $M_2=150 \G$ and the dashed ones to  
$M_2=300 \G$.}
\end{figure}

\begin{figure}
\epsfxsize=15.0 cm
\vspace{-2.5 cm}
\centerline{\epsfbox{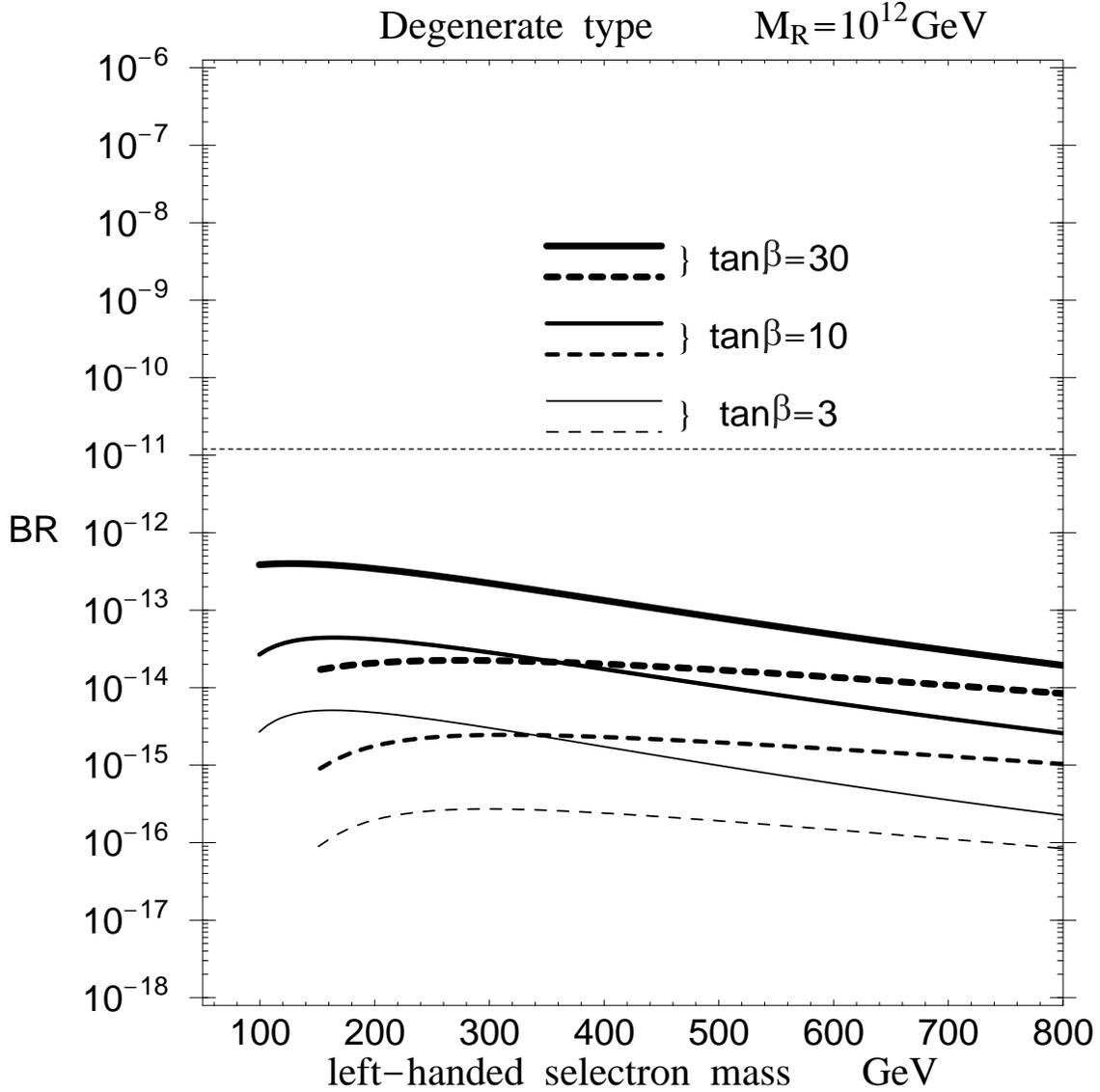}}
\caption{Predicted branching ratio BR$(\mu \rightarrow e \gamma)$
versus the left-handed selectron mass for $\tan\b=3,\ 10,\ 30$
in the case of the degenerate neutrino masses.
Here $M_R=10^{12}\G$ and $U_{e3}=0.2$ are taken.
The solid curves correspond to  $M_2=150 \G$ and the dashed ones to  
$M_2=300 \G$.}
\end{figure}

\begin{figure}
\epsfxsize=15.0 cm
\vspace{-2.5 cm}
\centerline{\epsfbox{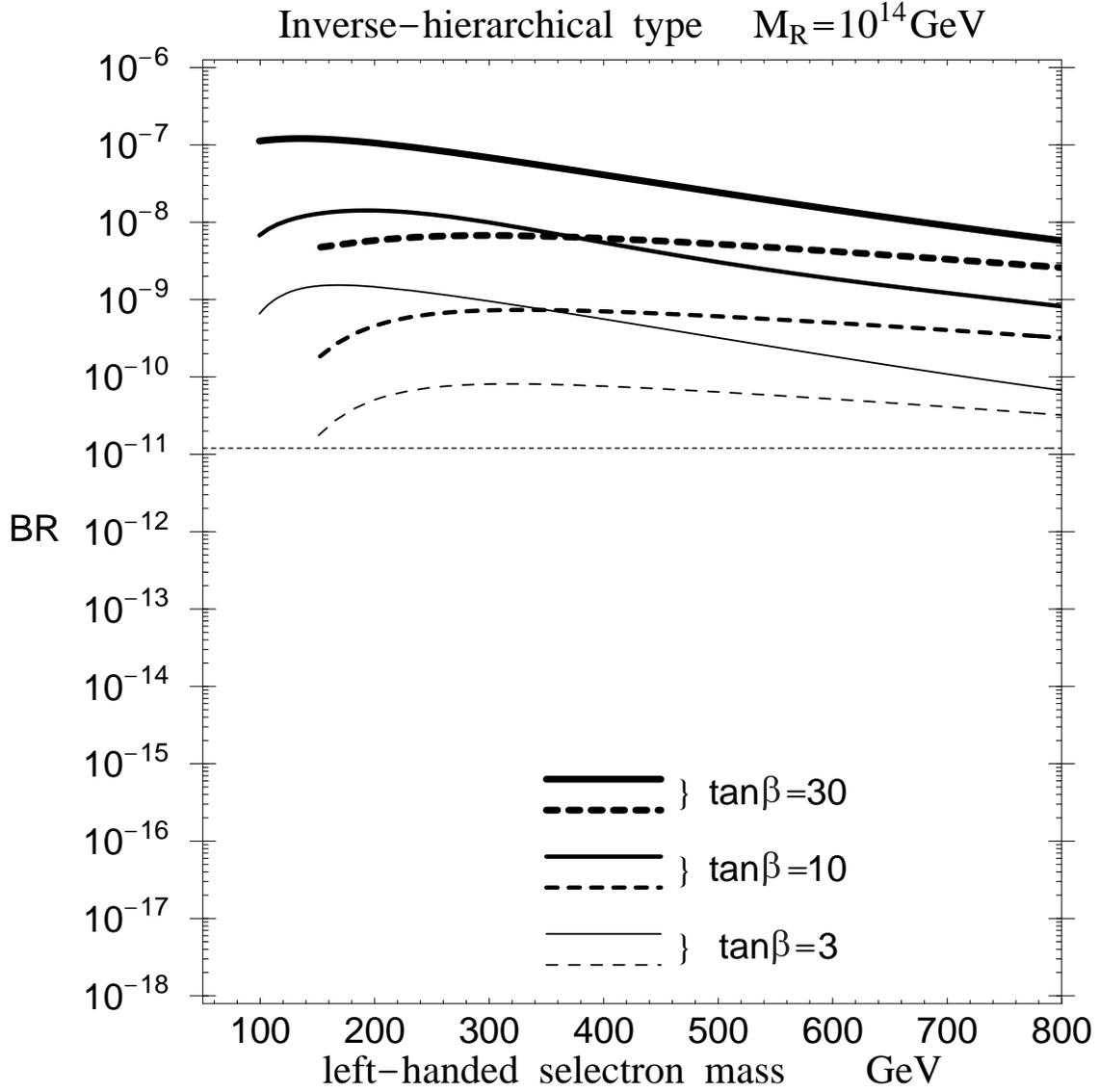}}
\caption{Predicted branching ratio BR$(\mu \rightarrow e \gamma)$
versus the left-handed selectron mass for $\tan\b=3,\ 10,\ 30$
in the case of the inverse-hierarchical neutrino masses.
Here $M_R=10^{14}\G$ and $U_{e3}=0.2$ are taken.
The solid curves correspond to  $M_2=150 \G$ and the dashed ones to  
$M_2=300 \G$.}
\end{figure}

\begin{figure}
\epsfxsize=15.0 cm
\vspace{-2.5 cm}
\centerline{\epsfbox{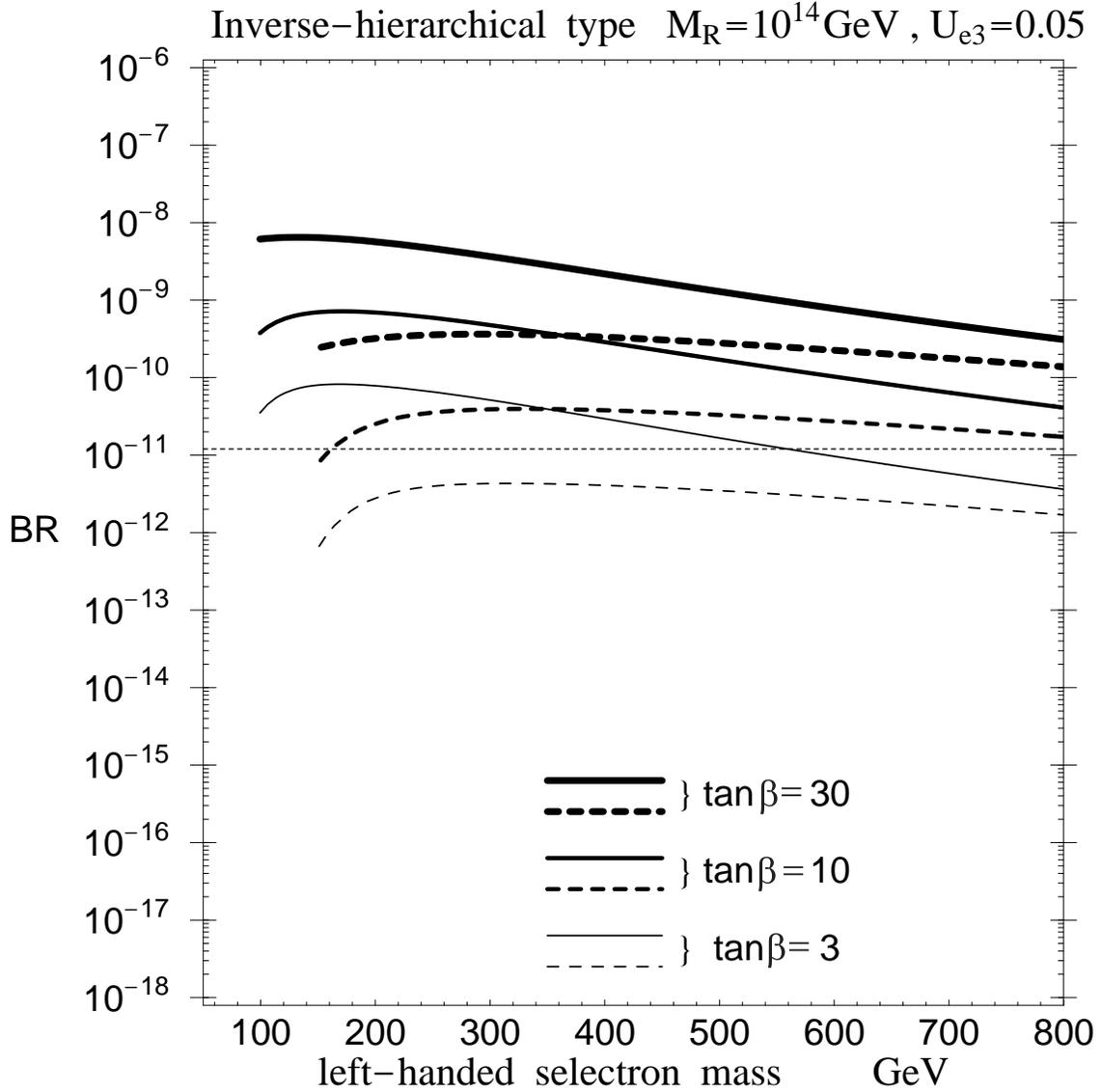}}
\caption{Predicted branching ratio BR$(\mu \rightarrow e \gamma)$
versus the left-handed selectron mass for $\tan\b=3,\ 10,\ 30$
in the case of the inverse-hierarchical neutrino masses.
Here $M_R=10^{14}\G$ and $U_{e3}=0.05$ are taken.
The solid curves correspond to  $M_2=150 \G$ and the dashed ones to  
$M_2=300 \G$.}
\end{figure}

\end{document}